\documentclass[pre,preprintnumbers,amsmath,amssymb,showpacs,
nofootinbib,floatfix]{revtex4}

\usepackage{graphicx,bm}

\makeatletter
\def\graphicscale{\twocolumn@sw{0.33}{0.4}}
\def\graphicthreescale{\twocolumn@sw{0.33}{0.4}}

\begin{document}

\title{On sampling and parametrization of discrete frequency distributions}

 \author{Paolo Rossi}
   \affiliation{Dipartimento di Fisica dell'Universit\`a di Pisa and
   I.N.F.N., Sezione di Pisa, Largo Bruno Pontecorvo 2, I-56127 Pisa,
   Italy} \date{October 3, 2012}

\begin{abstract}
The general relationship between an arbitrary frequency distribution and the expectation value of the frequency distributions of its samples is esablished. A set of combinations of expectation values whose value does not in general depend on the size of the sample is constructed. Distribution functions such that the distribution of the  expectation values of their samples is invariant in form are found and studied. The conditions under which the scaling limit of such distributions may exist are described.

\end{abstract}


\maketitle

%64.60.fd       General theory of critical region behavior
%64.60.F-       Equilibrium properties near critical points, critical exponents 
%67.25.dj       Superfluid transition and critical phenomena
%05.10.Cc       Renormalization group methods 
%05.70.Jk       Critical phenomena      

% ========================= BODY =========================
%\narrowtext

\section{Introduction}

In many interesting physical, biological and social phenomena, whenever no intrinsic scale for the relevant variables is present, the emergence of "scaling laws" is phenomenologically observed~\cite{Newman}.
However, strictly speaking, a power law is not a proper way of fitting empirical data, since no choice of the exponent can keep the higher moments of a power law distribution from diverging, while every phenomenological distribution leads to finite values for all moments.

This is not just a technicality: it is rather a reflection of the fact that the long tail of a power law distribution is in practice cutoffed by the existence of some "hidden" scale, irrelevant in the scaling region, but eventually forcing some upper limit on the variables describing the system.
It would therefore be convenient to be able to parametrize the data by means of more regular distribution functions, sufficiently damped for very large values of the variables, but at the same time admitting power law distributions as regular limits when some control parameter implementing the cutoff is sent to its limiting value.

A second, but not unrelated issue has to do with the effects of sampling, which may be non trivial even when we restrict our attention to the expectation values of the sampled variables.
On average sampling does not affect the distributions of individual objects belonging to different kinds, but when we consider frequency distributions (that is the number of kinds that are represented $k$ times in a given
population) we cannot in general expect that the frequency distribution in the samples be the same as in the 
original population, even after averaging on many different samples, basically because the cutoff induced by 
sampling acts differently (and in general nontrivially) at different scales.

Our purpose is therefore threefold. First we want to explore the general relationship existing between some arbitrary frequency distribution and the expectation value of the frequency distributions of its samples.

Moreover we want to study classes of distributions whose samples preserve (at least approximately) the functional 
dependence on the parameters present in the original distribution, establishing the connection between the 
(a priori unknown) values of the parameters of the distribution and the (empirically measured) parameters of the 
sample distributions.

Finally we want to study the scaling limit of these distributions (when it exists), in order to explore the possibility of their use for the phenomenological description of systems that are theoretically expected to show scaling in the limit when all empirical cutoffs are going to disappear.

In Section II we establish the notation and the general framework of our analysis, finding a rather explicit mathematical relationship between the generating function of the expectation values of the sample distributions and the generating function of the original distribution.

In Section III we construct an infinite set of combinations ("invariant moments") of expectation values whose value does not depend on the size of the sample. 

Evaluating the invariant moments on the samples leads therefore to the 
possibility of performing a formal reconstruction of the original distribution.

In Section IV we briefly discuss (as a corollary of the results presented in Sec.III) the issue of correlation between samples, which must be checked against its theoretical value as an important test of randomness in sampling.

In Section V we analyze a class of  distributions (the so-called negative binomial distributions) admitting a scaling limit and enjoying the property that the distribution of expectation values of the samples has the same mathematical form as the original distribution. We also compute in a closed form the values ot the invariant moments for these distributions.

In Section VI we consider two relevant generalizations, finding a quite general class of distributions sharing the property of invariance in form in the case of sampling, and describing the properties of some distributions that are not invariant in form for arbitrary values of the parameters but nevertheless admit a scaling limit.

Finally in Section VII we analyze  the scaling limit itself and discuss the conditions under which one may expect this limit to be a sensible description of the original system.

\section{The general framework}

We are considering a set of $N$ objects (``individuals'') belonging to $S$ different kinds (``species''), and we assume that the set contains $N_a$ objects
of the $a$-th kind, subject to the constraint
$$\sum_{a=1}^S N_a = N.$$

A sample is a set of $n$ objects, containing $n_a$ objects of the $a$-th kind, subject to the constraint 
$$\sum_{a=1}^S n_a = n.$$

The probability $P_{\{n_a\}}$ of extracting a specific sample $\{n_a\}$ from a given set $\{N_a\}$ is easily obtained and it takes the form
$$P_{\{n_a\}} = {N \choose n}^{-1}\prod_{a=1}^S {N_a \choose n_a}.$$

We can easily compute the relevant expectation values, obtaining in particular
$$<n_a> = N_a {n \over N}  = n\,p_a,$$
$$<n_a^2> = N_a (N_a-1) \Bigl({n(n-1) \over N(N-1)}\Bigr) + N_a {n \over N}
= {N \over N-1}n (n-1) p_a^2 +\bigl(1- {n-1 \over N-1} \bigr) n  p_a,$$
where $p_a \equiv {N_a/N}$ is the probability of extracting an object of the $a$-th kind in a single extraction.

It may be useful to consider also the limit of small samples $n_a << N_a$. In this limit the probability of a specific sample is well approximated by the expression
$$P_{\{n_a\}} = n! \prod_{a=1}^S {1 \over n_a!} (p_a)^{n_a}.$$

Notice that the condition  $<n_a> = n p_a$ is preserved by this approximation, while
$$<n_a^2> \rightarrow n(n-1) p_a^2 + n p_a.$$

A frequency distribution is a set of values $\{N_k\}$, where $N_k$ is the number of kinds such that for each of them there are $k$ objects in the original set. According to the definition, the following conditions must be satisfied:
$$\sum_{k=1}^N N_k = S, \qquad \qquad \qquad \sum_{k=1}^N k\,N_k = N.$$

The frequency distribution of a sample is a set of values $\{n_l\}$, satisfying the conditions
$$\sum_{l=0}^n n_l = S, \qquad \qquad \qquad \sum_{l=1}^n l\,n_l = n.$$

Notice that the frequency distribution of a sample includes the value $n_0$, 
corresponding to the number of kinds, present in the original set, which are not represented in the sample.

It is in principle possible to compute the probability of any sample distribution $\{n_l\}$ as a function of a given set $\{N_k\}$. To this purpose it is convenient to define the intermediate variables $N_{kl}$, representing the (random) number of kinds characterized by $k$ objects in the original set and by $l$ ($l \leq k$) objects in the sample. The variables $N_{kl}$ are strongly constrained, since they must satisfy all the conditions: 
$$\sum_{l=0}^n N_{kl} = N_k, \qquad \qquad \qquad \sum_{k=1}^N N_{kl} = n_l.$$

The probability $P_{\{N_{kl}\}}$ of a specific configuration ${\{N_{kl}\}}$ follows from the general probability formula:
$$ P_{\{N_{kl}\}} ={N \choose n}^{-1} \prod_{k=1}^N \biggl[N_k! \prod_{l=0}^k {1 \over N_{kl}!} {k \choose l}^{N_{kl}} \biggr],$$
subject to the constraint $\sum_{l=0}^n N_{kl} = N_k$.

The probability of finding a frequency distribution $\{n_l\}$ in a sample is then obtained by summing the probabilities $P_{\{N_{kl}\}}$ over all configurations  satisfiying the constraint $\sum_{k=1}^N N_{kl} = n_l$.

It may be convenient to define a multivariable generating function for the probability of the frequency distributions $P_{\{n_l\}}$:
$$E(x;\{t_l\}) \equiv  \sum_n  {N \choose n} x^n \sum_{\{n_l\}} P_{\{n_l\}} \prod_l t_l^{n_l}= \sum_n  {N \choose n} x^n \sum_{\{N_{kl}\}} P_{\{N_{kl}\}} \prod_l t_l^{\sum_k N_{kl}}.$$

By applying the explicit expression of $P_{\{N_{kl}\}}$ and all the constraints one may obtain the relationships:
$$E(x; \{t_l\}) =  \sum_{\{N_{kl}\}} \prod_k \Bigl({N_k! \over \prod_l {N_{kl}}!} \prod_l \Bigl[{k \choose l} t_lx^l\Bigr]^{N_{kl}}  \Bigr) = \prod_k \Bigl[ \sum_l {k \choose l}t_lx^l \Bigr]^{N_k}.$$

In practice it does not seem to be possible to obtain simple closed formulas for $ P_{\{n_l\}}$. However we shall be interested only in the expectation values $<n_l>$ of the frequency distribution, and these can be computed rather explicitly starting from the above expressions and from the relationship
$$<n_l> = \sum _{k=1}^N <N_{kl}> =\sum_{k=1}^N \sum_{\{N_{jm}\}}  N_{kl} P_{\{N_{jm}\}}.$$

Straightforward manipulations lead to the results
$$<N_{kl}> = N_k {{k \choose l}{N-k \choose n-l} \over {N \choose n}}, \qquad \qquad <n_l> = {\sum_{k=1}^N N_k {k \choose l}{N-k \choose n-l}  \over {N \choose n}}.$$

It is easy to check that the following relationships are satisfied:
$$\sum_{l=0}^n <n_l> = \sum_{k=1}^N N_k = S, \qquad \qquad
\sum_{l=0}^n l <n_l> = \Bigl(\sum_{k=1}^N k N_k \Bigr){n \over N} = n.$$

In order to fully appreciate the relevance of considerations based on the expectation values we must evaluate the weight of the fluctuations.  Taking second derivatives  of the generating function $E(x; t_i)$ one may obtain the expression:
$$<n_l^2>-<n_l>^2 = \sum_{k,k'} N_k N_{k'}  {k \choose l} {k' \choose l} \Biggl[{ {N-k-k' \choose n-2 l} \over {N \choose n}}-{ {N-k \choose n-l} \over {N \choose n}}{ {N-k' \choose n-l} \over {N \choose n}} \Biggr] +\sum_k N_k  {k \choose l}  \Biggl[{ {N-k \choose n- l} \over {N \choose n}}- {k \choose l} { {N-2 k \choose n-2 l} \over {N \choose n}} \Biggr].$$

A very important limit of the above result may be obtained when considering the (rather typical) case $k,l << N,n$. In this limit
$$<n_l> = \sum_{k=1}^N N_k {k \choose l} \bigl({n \over N}\bigr)^l \bigl(1-{n \over N} \bigr)^{k-l} \equiv  \sum_{k=1}^N N_k  P_{kl},$$
where the definition of $P_{kl}$ follows from the above equation, and one may check that the conditions on $\sum_{l=0}^n <n_l>$ and on $\sum_{l=0}^n l <n_l>$ are still satisfied.

We can also estimate the behavior of  fluctuations when $k,l << N,n$:
$$<n_l^2> - <n_l>^2 = \sum_k  N_k \Bigl[P_{kl}-P_{kl}^2\Bigr] -{N n \over N-n} \Bigl[\sum_k \Bigl({l \over n}-{k \over N} \Bigr) N_k P_{kl} \Bigr]^2.$$

It is easy to check that the above expression is always smaller than $<n_l>$ and as a consequence fluctuations become irrelevant for sufficiently large values of $<n_l>$.

In the same limit we may derive a very important relationship between the generating function of the original frequency distribution and the generating function of the expectation values of its samples.
Let us start by defining
$$F(t) \equiv \sum_{k=1}^N N_k t^k, \qquad \qquad f(t) \equiv \sum_{l=0}^n <n_l> t^l,$$
and let's notice that the above derived expression of $<n_l>$, when replaced in the definition of $f(t)$, after exchanging the order of summations and performing a summation on the index $l$ leads to
$$f(t) = \sum_{k=1}^N N_k (1-{n \over N}+{n \over N}t)^k = F(1-{n \over N}+{n \over N}t).$$
Even more conspicuously, by introducing a new variable $z$ and defining 
$$G(z) \equiv F(1-{z \over N}), \qquad \qquad \qquad g(z) \equiv f(1-{z \over n}),$$
we obtain the relationship
$$g(z) = G(z).$$

Defining $\gamma(z) \equiv G(z)-G(0)$ and recalling that $G(N) = 0$ and $G(n) = f(0)$ we then easily obtain
$$F(t) = \gamma \bigl[N(1-t)\bigr] -\gamma(N),$$
$$f(t) - f(0) = \gamma \bigl[n(1-t)\bigr] -\gamma(n).$$     

As a consequence, whenever the (size-independent) function $\gamma(z)$ can be cast into a form exhibiting no explicit parametric dependence on $N$, the expectation values  $<n_l>$  can be obtained from $N_k$ simply by the replacement $N \rightarrow n$.  

\section{Invariant expectation values}

It is very important to be able to define a set of expectation values that are independent of the size of the sample, and therefore may reflect very directly the properties of the original frequency distribution.

Let's consider the combinations 
$$\sum_{l=p}^n <n_l> {l \choose p} = {N \choose n}^{-1} \sum_{k=p}^N \sum_{l=p}^k N_k {l \choose p} {k \choose l} {N-k \choose n-l},$$
and exploit the fact that 
$${l \choose p} {k \choose l} = {k \choose p}{k-p \choose l-p}$$
and
$$\sum_{l=p}^k {k-p \choose l-p}{N-k \choose n-l} = {N-p \choose n-p} = {N \choose n} {n \choose p} {N \choose p}^{-1}$$
to prove that
$${n \choose p}^{-1} \sum_{l=p}^n <n_l> {l \choose p} = {N \choose p}^{-1} \sum_{k=p}^N N_k {k \choose p}.$$

Hence the following equations hold for all $p\leq n$:
$$<m_p^{(n)}> \equiv <{(n-p)! \over n!}\sum_{l=p}^n n_l {l \choose p}> = {(N-p)! \over N!} \sum_{k=p}^N N_k {k \choose p} \equiv M_p.$$

The expectation values of the ``moments'' $m_p^{(n)}$ evaluated for samples of arbitrary size $n$ coincide with the ``moments'' $M_p$ of the original frequency distribution, as long as $p\leq n$.

Notice that in the limit $k,l << N,n$ the definition of $m_p^{(n)}$ simplifies to
$$m_p^{(n)} \rightarrow {1 \over n^p}\sum_{l=p}^n n_l {l \choose p},$$
and the fact that $<m_p^{(n)}>$ are independent of the size of the sample then follows as a trivial consequence of the relationship $g(z)= G(z)$, holding in the same limit and allowing the interpretation of $G(z)$ as the generating function of the invariant moments.
\smallskip
It is worth analyzing the explicit expressions of the first few invariant moments:
$$M_0 \equiv \sum_{k=1}^N N_k = \sum_{l=0}^n <n_l> = S,$$
$$M_1 \equiv {1 \over N} \sum_{k=1}^N  k N_k = {1 \over n} \sum_{l=0}^n l <n_l> = 1,$$
$$2 M_2 \rightarrow {1 \over N^2} \sum_{k=1}^N  k(k-1) N_k = {1 \over n^2} \sum_{l=0}^n l(l-1) <n_l>,$$
hence
$$2 M_2 \equiv {1 \over \alpha} = \sum_{a=1}^S \bigl({N_a \over N}\bigr)^2 -{1 \over N} = \sum_{a=1}^S <\bigl({n_a \over n}\bigr)^2> -{1 \over n},$$
where we introduced the notation $\alpha = 1/(2 M_2)$ in order to make correspondence with the literature.

\section{Correlation between samples}
\bigskip
An important test of randomness in sampling is offered by the measure of the correlation between two different samples. Let's consider two random samples, characterized by the sets of values $\{n_a\}$ and $\{m_a\}$ and by their sizes $n$ and $m$. The index $a$ labels different kinds, as in Section II.

The correlation between the two samples is
$$C = {\sum_{a=1}^S n_a m_a \over \sqrt{\sum_{a=1}^S n_a^2}\sqrt{\sum_{a=1}^S m_a^2}}.$$

Replacing $n_a$ and $n_a^2$ with their expectation values, computed in Section II,
we then easily  obtain (in the large $N$ limit)
$$\sum_{a=1}^S <n_a> <m_a> = n m \sum_{a=1}^S p_a^2,$$
$$\sum_{a=1}^S <n_a^2> = n^2 \bigl(\sum_{a=1}^S p_a^2 +{1 \over  n}-{1 \over N}\bigr), \qquad \qquad \sum_{a=1}^S <m_a^2> = m^2 \bigl(\sum_{a=1}^S p_a^2 +{1 \over m}-{1 \over N}\bigr).$$

By making use of the results presented at the end of the previous Section we can now express the expected value of the correlation between samples in the form
$$<C> = {{1 \over \alpha} + {1 \over N} \over \sqrt{{1 \over \alpha} + {1 \over n}} \sqrt {{1 \over \alpha} + {1 \over m}}}.$$

For samples of equal size $n$ we can represent the expected value of the correlation in the form:
$$<C> = {n \over \alpha+ n}{\alpha +N \over N}.$$

\section{A class of distributions and its properties}

For our purposes it is especially interesting to consider the class of negative binomial distributions~\cite{Hilbe}, which can be obtained starting from the generating function

$$ F_c (t) = {N \over x}{(1-x)^{1-c} \over c} \bigl[1-(1-x t)^c \bigr] = \sum_{k=1}^\infty {N \over x}{(1-x)^{1-c} \over \Gamma(1-c)}{\Gamma(k-c) \over k!} x^k,$$
where $0<x<1$ and the parameter c is assumed to vary in the range $0 \leq c <1$.

The asymptotic behaviour of the distribution for large $k$ is easily obtained by observing that in this limit
$${\Gamma (k-c) \over k!} \rightarrow {1 \over k^{1+c}}, \qquad \qquad
N_k \rightarrow {N \over x}{(1-x)^{1-c} \over \Gamma(1-c)}{x^k \over k^{1+c}}.$$

We can now compute the generating function for the expectation values of the samples according to the general rule previously discussed, and obtain
$$f_c(t) =f_c(0)+{n \over y}{(1-y)^{1-c} \over c} \bigl[1-(1-y t)^c \bigr],$$
where we have defined 
$$y = {{n \over N}x \over 1-x+{n \over N}x}.$$

The distribution of the samples has the same form as the original distribution, once the replacements $N \rightarrow n$ and $x \rightarrow y$ have been performed, and therefore we obtain the asymptotic behaviour
$$n_l \rightarrow {n \over y}{(1-y)^{1-c} \over \Gamma(1-c)}{y^k \over k^{1+c}}.$$

It is possible to define a combination of parameters independent of the dimension of the sample:
$$\beta = N{1-x \over x} = n{1-y \over y},$$
and it is useful to represent $x$ and $y$ in a form showing explicitly their dependence on the dimension of the sample and on the invariant parameter $\beta$:
$$x = {N \over \beta+N}, \qquad \qquad \qquad y = {n \over \beta+n}.$$

It is now possible to evaluate the invariant moments from the expression
$$\gamma_c(z) \equiv G_c(z)-G_c(0) = {\beta \over c}\Bigl[ 1-\Bigl(1+{ z \over \beta}\Bigr)^c\Bigr],$$
showing no explicit parametric dependence on $N$, as expected; we therefore obtain (for $p \neq 0$)
$$M_p = {\beta^{1-p} \over \Gamma(1-c)}  {\Gamma (p-c) \over p!} \rightarrow {1 \over \Gamma(1-c)} {\beta^{1-p} \over  p^{1+c}}, \qquad \qquad \alpha \equiv {1 \over 2 M_2} = {\beta \over 1-c}.$$

For completeness let's observe that 
$$S = G_c(0) \equiv F_c(1) \equiv f_c(1) = {\beta \over c} \bigl[(1-x)^{-c}-1\bigr], \qquad s = f_c(1) - f_c(0) = {\beta \over c} \bigl[(1-y)^{-c}-1\bigr],$$
and as a consequence  the expected number of kinds not appearing in a given sample is
$$n_0 \equiv f_c(0) = S-s = {\beta \over c} \bigl[(1-x)^{-c}-(1-y)^{-c}\bigr].$$

The limit of the above results when $c \rightarrow 0$ is smooth, and it corresponds to Fisher distribution~\cite{Fisher}, such that
$$F_0(t) = - \beta \ln (1-x t), \qquad \qquad \qquad N_k = \beta {x^k \over k},$$
and
$$f_0(t) = \beta  \ln \bigl({\beta+N \over \beta+ n}\bigr) -\beta \ln (1-y t), \qquad \qquad n_l = \beta {y^l \over l}.$$

The generating function of the invariant moments is obtained from
$$\gamma_0(z) =  - \beta \ln \bigl(1+ {z \over \beta} \bigr),$$
and as a consequence the invariant moments ($p \neq 0$) are exactly $M_p = \beta^{1-p} / p$.

Notice in particular the relationship $\beta = \alpha$, peculiar to Fisher distribution.

\section{More general distributions}

Negative binomial distributions are only a special instance of a much wider class of distributions whose generating function can be represented in the form
$$F(t) = {\beta \over \varphi'(1)} \Bigl[ \varphi\bigl({1 \over 1-x }\bigr)- \varphi \bigl({1- x t \over 1-x }\bigr) \Bigr],
$$
where $\varphi$ is an arbitrary function of its argument (subject only to the constraints deriving from the positivity of all $N_k$) and $x$ is related to $N$ and $\beta$ as in the previous Section.

We can easily evaluate the generating function of the invariant moments, finding
$$\gamma(z) = {\beta \over \varphi'(1)} \Bigl[ \varphi (1)- \varphi \bigl(1+ {z \over \beta}\bigr) \Bigr],$$
and therefore we obtain
$$f(t) - f(0) = {\beta \over \varphi'(1)} \Bigl[ \varphi\bigl({1 \over 1-y }\bigr)- \varphi \bigl({1- y t \over 1-y }\bigr) 
\Bigr],$$
showing explicitly that the distribution of expectation values in the samples preserves the the form of the original frequency distribution.

Ii is worth noticing that, whenever $\varphi (u) \rightarrow  u^c $ in the limit $u \rightarrow 1$, the asymptotic 
behaviour of the distribution for large values of $k$ will be the same as in Section V, independent of the detailed 
behaviour exhibited for small $k$.
\bigskip
It is also worth considering the class of generating functions admitting the representation
$$F(t) = {N \over x}{\psi (x t) \over \psi'(x)}.$$

In the general case the sample distribution functions will not preserve strictly the form of the original distribution, but once again, admitting that, in the limit $t \rightarrow 1$, 
 $\psi (1-\xi) \rightarrow  \xi^c $, one can show that the relevant asymptotic behaviours are
$$N_k \rightarrow A(x) \Bigl({N \over x} \Bigr)^c \beta^{1-c} {x^k \over k^{1+c}}, \qquad \qquad  n_l \rightarrow A(x) \Bigl({n \over y} \Bigr)^c \beta^{1-c} {y^l \over l^{1+c}},$$
$$M_p \rightarrow A(x) {\beta^{1-p} \over p^{1+c}},$$
where we have introduced the invariant coefficient 
$$A(x) = {1 \over \Gamma(1-c)}{c\,(1-x)^{c-1} \over \psi'(x)},$$
and it is possible to show that, independent of the detailed form of $\psi$, the limit of $A(x)$ when 
$x \rightarrow 1$ must always be equal to $1/\Gamma(1-c)$.

\section{The scaling limit}

Let's now consider very large systems, and assume that we can gather information only through the sampling of
$n$ objects belonging to the system, with $n$ large but not necessarily comparable to $N$.

The analysis of the invariant moments may then allow us to check the applicability of a phenomenological description of the samples based on some distribution falling into the classes discussed in the previous Sections.
In the case of a positive response to the check it is then possible to find numerical estimates of the parameter $\beta$ and of the exponent $c$ (as well as of the coefficient $A(x)$, if present).
Such estimates are clearly meaningful only if $\beta$ does not turn out to be significantly greater than $n$.

Under these assumptions, we can infer a description of the original system, and in the case $N >> n$ such a description will correspond to computing the limit $x \rightarrow 1$ of the previous results.
As a consequence, at least for observable (i.e. not too large) values of $k$, the original distribution is expected to be well described by the scaling form
$$N_k \rightarrow  {N^c \beta^{1-c} \over \Gamma(1-c)}{1 \over k^{1+c}}.$$

\begin{acknowledgments}

I am indebted to Sergio Caracciolo for an important observation that led to reconsidering and clarifying the r\^ole of fluctuations. I am also indebted to Steve Shore and Ettore Vicari for critical reading of the manuscript.

\end{acknowledgments}

\end{document}